\documentclass{iopart}
\usepackage{times}
\begin{document}

\title{Insights from simulations of star formation}

\author{Richard B Larson}

\address{Department of Astronomy, Yale University, Box 208101, New Haven,
         CT 06520-8101, USA}
\ead{richard.larson@yale.edu}

\bigskip\bigskip
\noindent\hskip-1pt{\small
{\sl `The purpose of computing is insight, not numbers.'}\\
Richard W. Hamming, in Numerical Methods for\\
Scientists and Engineers (1962)

\bigskip
\noindent\hskip-1pt
{\sl `There are more things in heaven and earth, Horatio,\\
than are dreamt of in your philosophy.'}~~William\\
Shakespeare, in Hamlet, Prince of Denmark (1604)}
\bigskip

\begin{abstract}

   Although the basic physics of star formation is classical, numerical simulations have yielded essential insights into how stars form.  They show that star formation is a highly nonuniform runaway process characterized by the emergence of nearly singular peaks in density, followed by the accretional growth of embryo stars that form at these density peaks.  Circumstellar disks often form from the gas being accreted by the forming stars, and accretion from these disks may be episodic, driven by gravitational instabilities or by protostellar interactions.  Star-forming clouds typically develop filamentary structures, which may, along with the thermal physics, play an important role in the origin of stellar masses because of the sensitivity of filament fragmentation to temperature variations.  Simulations of the formation of star clusters show that the most massive stars form by continuing accretion in the dense cluster cores, and this again is a runaway process that couples star formation and cluster formation.  Star-forming clouds also tend to develop hierarchical structures, and smaller groups of forming objects tend to merge into progressively larger ones, a generic feature of self-gravitating systems that is common to star formation and galaxy formation.  Because of the large range of scales and the complex dynamics involved, analytic models cannot adequately describe many aspects of star formation, and detailed numerical simulations are needed to advance our understanding of the subject.

\end{abstract}

\section{Introduction}

   Astrophysics strives to understand astronomical phenomena using the laws of physics and the tools of mathematics.  Newton famously showed in his Principia Mathematica how the techniques of calculus could be used to solve the two-body problem and thus derive from physical principles Kepler's laws for the motions of the planets around the Sun.  However, Newton also struggled long and ultimately unsuccessfully to explain equally well the motion of the Moon around the Earth as both orbit the Sun.  We now understand that the motion of the Moon, as an example of the three-body problem, has no analytic solution and can only be solved approximately.  But thanks to modern computers we can now calculate the motion of the Moon with high accuracy by integrating numerically the equations of motion of all of the bodies involved, including other planets as well as the Sun, Earth, and Moon.  The numerical techniques used for this purpose approximate the motions of these bodies with a series of small but finite steps, at each of which all of the forces are recalculated and the motions of the bodies are extrapolated to the next step.  Sophisticated algorithms and powerful computers enable us to calculate orbits with great precision over long periods of time, and the power of these techniques is routinely demonstrated by the impressive precision with which we can navigate spacecraft to any desired destination in the solar system. 

   Numerical methods utilizing finite space and time steps have been applied in many areas of science over the past half-century, and they have expanded enormously our ability to model and understand natural phenomena.  Detailed numerical simulations have allowed many new problems to be solved and many old ones to be advanced to a higher level of understanding.  But perhaps the most important contribution of numerical techniques to science has been that they have often discovered new phenomena or revealed unexpected results whose importance had not previously been recognized.  In doing so, they have greatly expanded our ideas about what can happen in complex systems for which no analytic solutions exist and the laws of physics may allow many outcomes; in effect, they have provided a powerful exploratory tool that can supplement our limited imaginations and provide new insights into how nature works.  In astronomy, a classic and elegant example of how numerical techniques can reveal an unexpected richness of phenomena was provided by the work of Toomre and Toomre (1972), who used numerical integration of the restricted three-body problem (two massive bodies and one massless one) to model tidal interactions between galaxies; the results were dramatic and showed immediately that many strikingly peculiar galaxies could be understood as gravitationally interacting systems.  This work launched the whole new field of study of galaxy interactions, a phenomenon whose importance had not previously been realized.

   Even systems governed by simple laws can quickly develop a level of complexity that surpasses our ability to form a simple mental picture or model, and in such cases computer simulations can often be used to gain understanding.  A common way in which complexity can emerge is via the chaotic behavior that characterizes many natural phenomena and makes them unpredictable, even in principle, over extended periods of time.  An example is provided by the three-body problem, in which the extreme sensitivity of the orbits to the initial conditions can cause them to diverge exponentially and make them impossible to predict over indefinite periods of time.  A three-body system generally decays eventually into a binary system and an escaper, but the time and manner of such a decay cannot in general be predicted a priori, and can only be studied statistically via extensive numerical experimentation.  Many astrophysical problems are much more complex than this because they involve fluid dynamics which has many more degrees of freedom and is susceptible to its own kind of chaotic behavior via the development of turbulence.  A familiar example of the unpredictability of fluid motions over long periods of time is provided by the fact that the weather cannot be predicted very far into the future.  Many astrophysical problems involve both gravity and fluid dynamics, and this gives them multiple ways to develop complex and chaotic behavior.

   All but the simplest problems in fluid dynamics can be solved only with numerical methods.  The techniques that were first used to solve problems in astrophysical fluid dynamics had their origins in the nuclear weapons programs of the cold war in the 1950s, where they were needed to model powerful explosions capable of creating extreme physical conditions.  Practitioners of this art later brought some of the techniques involved to other areas of science and engineering, and in the academic world they found some of their first scientific applications in astrophysics, where similar processes and conditions can often be important.  For example, many astrophysical problems involve radiative processes and fluid dynamics with shock fronts, and since both of these phenomena had been addressed numerically in the bomb codes, the techniques involved could be transferred fairly directly to astrophysical problems.

   One of the first astrophysical problems to be studied numerically, following the widespread introduction of computers and numerical techniques into the academic world in the 1960s, was the formation of stars by the gravitational collapse of interstellar gas clouds.  This problem involves both gravity and gas dynamics, and in general it can be solved only numerically, despite the fact that the basic physics involved is classical and well known; only a few of the simplest problems in star formation theory can be solved analytically.  Numerical simulations of star formation currently constitute a sizeable and growing research area, and much of the recent progress and ongoing debates in star formation theory have come from the results of numerical work.  The physics of star formation and some of the most important results of simulations were summarized in an earlier review in this journal (Larson 2003), and the present article supplements that review by focusing more on the overall picture emerging from the simulations and on the insights that they have provided about how stars form.

\section{Some challenges in star formation theory}

   Most of the visible matter in the universe is condensed into tiny, almost point-like objects, i.e.\ stars, with densities more than 30 orders of magnitude higher than the average density of the universe and more than 20 orders of magnitude higher than the densities of the interstellar clouds in which they form.  Yet, we know from observations of the cosmic background radiation that that the universe was once exceedingly uniform.  How did its visible matter become so highly condensed into these tiny objects?  The problem is not just one of understanding how the cosmic baryons became condensed into stars, because most of the baryons actually did not go into stars but have remained behind in a hot dilute intergalactic medium.  The problem is thus really one of understanding how the baryons became separated into these very different phases with such a huge contrast in density.

   Speculations about how stars form date back to Newton, who suggested in 1692 that gravity might cause matter uniformly spread throughout an infinite space to be collected into an infinite number of stars\footnote{In his first letter to Bentley, as quoted by Jeans (1929), Newton said ``It seems to me, that if the matter of our sun and planets, and all the matter of the universe, were evenly scattered throughout all the heavens, and every particle had an innate gravity towards all the rest, and the whole space throughout which this matter was scattered, was finite, the matter on the outside of this space would by its gravity tend towards all the matter on the inside, and by consequence fall down into the middle of the whole space, and there compose one great spherical mass.  But if the matter were evenly disposed throughout an infinite space, it could never convene into one mass; but some of it would convene into one mass and some into another, so as to make an infinite number of great masses, scattered great distances from one to another throughout all that infinite space.  And thus might the sun and fixed stars be formed, supposing the matter were of a lucid nature.''}.  We know now that Newton's speculation that stars are formed by gravity was indeed correct, and we also know that star formation has continued throughout most of cosmic history and is still going on in the dense molecular clouds that we see around us in the interstellar medium.  If the collapse of these molecular clouds into stars is not prevented or strongly resisted by forces that oppose gravity, the time that would be required for them to collapse under their gravity is less than 10 million years; the simplest ideas about star formation would then suggest that the observed clouds should all collapse completely into stars in a time not much longer than this.  But this clearly does not happen: only a small fraction of the matter in a molecular cloud actually ends up in stars, and our Galaxy still retains a considerable amount of interstellar gas even after more than 10 billion years of Galactic history.  It is therefore a challenge for theory to understand why the process is so inefficient on both cloud scales and cosmic scales.

   Another challenge is to understand what happens to the angular momentum and magnetic flux that are present in typical star-forming clouds; if these quantities were conserved during cloud collapse, centrifugal and magnetic forces would eventually overwhelm gravity and prevent stars from forming.  Evidently, angular momentum and magnetic flux are somehow efficiently removed or redistributed during star formation.  Assuming that these obstacles can be overcome, we still need to understand the typical properties of the stars and stellar systems that form.  The most fundamental property of a star is its mass; most stars have masses between about 0.01 and 100 times the mass of the Sun, and a typical stellar mass is a few tenths of a solar mass.  What determines this typical mass, which is a quantity of fundamental importance throughout astronomy and astrophysics?  A further crucial fact that needs to be understood is that many stars, including most of the more massive ones, occur in binary or multiple systems, whose evolution can play an important role in many astrophysical processes.  Most stars also form in larger groupings and clusters, so we need also to understand the formation of clusters of stars.  Finally, recent observations have shown that planetary systems are common, but the systems discovered so far are quite diverse, and most of them have little resemblance to our own Solar system.  How can we understand this great diversity of planetary systems, and is our own Solar system special in some way?  These questions all pose challenges for star formation theory, and answering them is a central goal of much current research.

\section{Insights from collapse calculations}

   The simplest problem in star formation theory is the free-fall collapse of a uniform pressure-free sphere; this problem can be solved analytically, and the result is that the sphere undergoes a runaway collapse to an infinite density in a finite time $t_{\rm ff} = (3\pi/32G\rho)^{1/2}$ (Spitzer 1978).  Any departure from this simple case, such as a nonuniform density or a finite pressure, produces a more complicated result that depends on the initial and boundary conditions and cannot be derived analytically, making numerical calculations necessary.  One of the first and still most important results of early numerical calculations of the collapse of a sphere with finite pressure was that if collapse can occur at all, it leads not just to a runaway increase in density but to a runaway increase in the density contrast between a growing central density peak and a surrounding envelope that merges with the surrounding medium and collapses relatively little while the central density approaches stellar density (Penston 1966, 1969b; Bodenheimer and Sweigart 1968; Larson 1969).  Although it might be expected that any initial density variation would be amplified during collapse because the free-fall time depends inversely on density, so that denser regions tend to collapse faster than less dense ones, it had not been anticipated before the calculations were done how dramatic this effect can be, leading to an enormous range in density all the way from stellar to interstellar in a collapsing cloud.  If the usual isothermal approximation is adopted for the early stages of collapse, the density profile approaches the power-law form $\rho \propto r^{-2}$, and if isothermality continues to hold, a central singularity in density is approached in only a little more than a free-fall time.

   The approach to a central density singularity is perhaps the essential mathematical feature of star formation, because stars are so small compared with interstellar distances as to be mathematically almost mass points.  Detailed collapse calculations show that gravity can indeed create condensed objects that are almost mass points, embedded in extended envelopes that merge with the surrounding medium.  A nearly singular density profile results because the central densest part of a collapsing cloud collapses faster than the less dense outer parts, causing the density gradient to become steeper and steeper near the center; the increase in the pressure gradient in this region then tends to limit the collapse rate as the central density continues to increase indefinitely.  The result is that pressure forces never become negligible compared with gravity, and pressure and gravity forces eventually approach a constant ratio; in the isothermal case this implies a density profile of the form $\rho \propto r^{-2}$.  This important result of collapse calculations was not predicted or anticipated before the calculations were done, but a singular isothermal sphere has since become a widely used approximation for modeling star-forming cloud cores (e.g., Shu 1977; Shu \etal 1987).  This is an example of how numerical calculations can aid imagination and provide new insights by yielding an unexpected result that is of fundamental importance for a problem that lacks an analytic solution.

   Once these numerical results had been obtained, it was realized that they were approaching a self-similar form that could be described by a similarity solution (Larson 1969; Penston 1969a).  In this similarity solution, the shape of the density profile remains invariant with time, approaching $\rho \propto r^{-2}$ at large radii, while the scale factors for radius and density vary with time in such a way that the $r^{-2}$ density profile becomes extended inward indefinitely to smaller and smaller radii.  This type of solution can be generalized to the case of a polytropic equation of state $P \propto \rho^{\gamma}$, for which the asymptotic density profile is $\rho \propto r^{-2/(2 - \gamma)}$ (Larson 1969; Ogino \etal 1999).  Collapse calculations started from different initial conditions tend to converge toward these similarity solutions, although they may never actually approach them very closely in real cases.  Further work (Hunter 1977; Whitworth and Summers 1985) has demonstrated the existence of a whole family of similarity solutions of which the Larson-Penston solution is a limiting case representing the fastest possible collapse that a partially pressure-supported cloud can undergo once the initial and boundary conditions are no longer important.  These similarity solutions have been widely used to approximate the limiting behavior of collapsing clouds because of their simplicity and generality, being independent of the initial and boundary conditions.

   The approach to a density singularity and the existence of a similarity solution describing it occur not only in spherical collapse but also in collapse with rotation (Norman \etal 1980; Narita \etal 1984; Matsumoto \etal 1997; Saigo and Hanawa 1998; Saigo \etal 2000) and with a magnetic field (Basu and Mouschovias 1994; Basu 1997; Nakamura \etal 1999).  In these cases, the collapse is non-spherical and occurs preferentially along the rotation axis or the field lines to create a flattened disk-like configuration, but it still produces a central density peak that evolves toward a singularity, and the radial density profile in the flattened envelope around it has the same power-law form as in the spherical case.  The calculations are much more challenging in this case than in the spherical case because great attention must be paid to adequate numerical resolution near the center, and this result became clear only after years of work; in fact, some of the earliest calculations of collapse with rotation (Larson 1972; Black and Bodenheimer 1976) had produced rings which now appear to be artifacts of insufficient numerical resolution near the center.  However, it is now clear that as long as collapse is not completely prevented by rotation or a magnetic field, isothermal collapse still produces an approach to a central density singularity.  This result suggests that the earliest stages of star formation occur in qualitatively the same way even with rotation or a magnetic field, as long as these effects do not prevent collapse.  The prediction that star formation begins with the development of a nearly singular peak in density contrasts with earlier more naive ideas whereby a prestellar gas clump was imagined to collapse as a unit and leave behind an evacuated region.  Instead, we now understand that when pressure is properly included, the result is the emergence of a near-singularity in density surrounded by a smoothly falling density profile that merges into the surrounding medium.

   A key implication of the prediction that star formation begins with the formation of a nearly singular peak in density is that stars begin their existence as tiny seed objects or stellar `embryos' that form at such density peaks and proceed to grow in mass by accreting matter from the surrounding envelope.  Stars are then predicted to gain most of their mass by accretion, so that the star formation process can be regarded as being largely an accretion process.  Because of this, much effort has gone into studying protostellar accretion and the interactions that can occur between accreting stars and their surrounding envelopes via the effects of radiation and outflows from the central star.  An important type of accretion problem to be discussed further below is the problem of accretion from circumstellar disks that form from infalling matter that has too much angular momentum to fall directly onto the accreting stars.

   In a large enough cloud, many smaller regions can be gravitationally bound and can collapse individually.  The number and typical mass of these collapsing regions can be estimated from the analysis first made by Jeans (1902) of the stability of an infinite uniform medium with finite pressure, which shows that there is a minimum size and mass for a region that can collapse under its self-gravity.  Real star-forming clouds are of course much more complex than the infinite uniform medium considered by Jeans, and the Jeans analysis was inconsistent in neglecting the overall collapse of the medium; nevertheless, much work has shown that the basic dimensional result of the Jeans analysis remains valid even in the presence of complicating effects such as non-spherical geometry, rotation, turbulence, and magnetic fields, as long as these effects do not completely dominate (see the review by Larson 2003).  In the densest parts of molecular clouds, the Jeans mass is of the order of one solar mass and therefore is much smaller than the typical cloud mass of many thousands of solar masses.  This suggests that large molecular clouds should form many stars, and this is confirmed by detailed three-dimensional collapse simulations which show that large clouds always fragment into many small clumps, typically in groups and clusters resembling the observed groupings of young stars in star-forming clouds (Klessen and Burkert 2001; Bonnell and Bate 2002; Bate \etal 2003; Bonnell \etal 2003).  Since each star is predicted to begin its existence as a nearly singular density peak, many such density peaks should form in a time not much longer than the free-fall time; this is indeed seen in the simulations, although with relatively low resolution in the 3-D case.  Star formation can thus perhaps be thought of as a precipitation process whereby many condensation nuclei precipitate out of a collapsing cloud and proceed to grow by accreting mass.

   The highly non-uniform nature of the star formation process shown by the simulations has fundamental implications for cloud fragmentation and stellar masses because it strongly limits the amount of fragmentation that can occur.  The influential concept of hierachical fragmentation proposed by Hoyle (1953) supposes that a cloud collapses nearly uniformly, so that all parts of the cloud undergo nearly the same increase in density.  The Jeans mass, which depends inversely on density, then decreases by a similar amount everywhere, allowing the cloud to fragment into progressively smaller units as it collapses.  Such a fragmentation process might in principle proceed until the increase of opacity at very high densities makes the isothermal approximation no longer valid (Low and Lynden-Bell 1976).  However, it is now clear that the hierarchical fragmentation process envisioned by Hoyle cannot occur, or at best can occur only in a much more limited way, because clouds do not collapse nearly uniformly but develop extreme density variations such that only a small fraction of the mass goes into the growing density peaks while most of it remains behind at lower density and cannot fragment.  Collapsing clumps may if they are rotating fragment into binary or multiple systems, but apart from this possibility, the amount of fragmentation that occurs during isothermal collapse is likely to be strongly limited by the initial conditions.  The number of bound clumps that form in many numerical simulations of collapsing and fragmenting clouds is in fact of the same order as the number of Jeans masses present initially, as will be discussed further in section 6. 

   The non-uniform nature of the star formation process also makes it understandable how star formation can be rapid yet inefficient.  Only a small fraction of the mass in a collapsing cloud goes into the accreting stellar embryos or protostars, and most of the mass remains behind in diffuse envelopes that merge into the surrounding medium.  Once a protostar has formed and begun to accrete, it can become a significant energy source and can begin to influence the surrounding envelope through feedback effects such as radiative heating, radiation pressure, outflows, and eventually ionization in the case of massive stars.  All of these effects can inhibit continuing accretion, and a combination of them may disperse most of the gas in a star-forming cloud before it has had time to condense into stars, severely limiting the efficiency of star formation.  Although these feedback effects are not yet quantitatively well understood, observations confirm that such phenomena do indeed occur and can disperse a significant fraction of the material in a star-forming cloud (e.g., Arce and Sargent 2006).  A low efficiency of star formation is therefore a natural consequence of the non-uniform nature of the star formation process.  The relatively low rate, or long timescale, for star formation in our Galaxy in comparison with the relevant dynamical timescales can then be understood as a consequence of its low efficiency, and it does not require, as was once thought, that the collapse of star-forming clouds be strongly resisted by magnetic fields or other effects.  Collapse can occur on a free-fall timescale and star formation can still be very inefficient.

\section{The role of rotation, disks, and jets}

   Simulations of star formation that include rotation or turbulence show that the gas being accreted by the forming stars often forms disks around them because it has too much angular momentum to fall directly onto these stars.  For example, the simulation by Bate \etal (2002b, 2003) of the formation of a small cluster of stars shows clearly the formation of many circumstellar disks, despite the fact that the dynamics of the system is complex and chaotic and protostellar interactions are frequent.  The disks that form are often disrupted by encounters, but new disks soon form from gas that continues to fall inward and swirl around the forming stars.  These results are consistent with the extensive observational evidence indicating that disks are a common feature of star formation (Hartmann 1998).  Additional evidence is provided by the fact that planetary systems have recently been found to be common, and like our own Solar System, they almost certainly originated in disks of some kind.

   Since most of the gas that falls into a circumstellar disk is believed to be accreted eventually by the central star, much work has gone into trying to understand accretion from protostellar disks, but this remains a difficult and poorly understood subject.  The original idea behind the standard disk models often used was that turbulence might provide an effective source of viscosity to transport angular momentum outward and hence drive an accretion flow, but simulations of turbulent disks have generally not supported this idea, and some have even shown that angular momentum is transported {\it inward,} i.e., in the wrong direction to drive an accretion flow (Stone \etal 2000). Fluid turbulence therefore seems unpromising as a mechanism to drive disk accretion, although if magnetic fields are present, MHD turbulence may provide a source of viscosity that can drive accretion in regions having a sufficient degree of ionization (Balbus and Hawley 1998; Stone \etal 2000).  Another possibility that has been studied extensively with numerical simulations is the effect of gravitational instabilities in disks; the associated gravitational torques can play an important role in transporting angular momentum outward in a massive protostellar disk if radiative cooling is important (Bodenheimer 1995; Stone \etal 2000; Lodato and Rice 2005).  Instabilities of moderate amplitude can produce transient spiral density fluctuations whose associated torques drive an accretion flow (Larson 1984; Gammie 2001), while instabilities of larger amplitude may create bound clumps, either transient or permanent, which produce strong disturbances that rapidly restructure the disk (Vorobyov and Basu 2006).  These clumps may even lose enough angular momentum by interaction with the disk that they rapidly spiral inward and produce bursts of accretion onto the central star; this `burst mode' of accretion could even account for most of the mass gained by the central star (Vorobyov and Basu 2006).  Further work is needed to confirm this striking result of the simulations, which implies that accretion may not be a steady process but may instead be episodic and occur mostly in bursts.

   Evidence that protostellar accretion is at least partly episodic is provided by the FU Orionis flareups that have been observed in some newly formed stars, and that are believed to be powered by episodes of exceptionally rapid accretion (Hartmann and Kenyon 1996).  The infall of massive clumps from a disk might be one way to produce these flareups, as suggested by Vorobyov and Basu (2006).  Another possibility is that tidal disturbances due to interactions with other stars in a dense environment, for example interactions with a binary companion that makes periodic close passages, might perturb a circumstellar disk sufficiently to trigger episodes of rapid accretion (Bonnell and Bastien 1992).  Other explanations have also been offered for the FU Orionis phenomenon (Hartmann and Kenyon 1996), but these two gravitational mechanisms are probably the simplest ones in terms of the physics involved.  Both were suggested by the results of numerical simulations, which showed that even classical gravitational physics can have dramatic and unexpected consequences in a disk.  The effects of tidal interactions on protostellar disks resemble the effects of such interactions on spiral galaxies, which is not surprising given the similar physics involved.  Some of the insights that have been gained from the study of galaxy interactions may therefore apply also to the effects of interactions on protostellar disks.  For example, in the galactic case it is clear that tidal interactions can cause gas to fall rapidly toward the center of a galaxy and produce strong starburst and AGN activity; analogous events in protostellar disks might cause bursts of accretion onto the central star, resulting in flareups and possibly jet production (see below).

   Simulations have thus shown that the dynamics of disks can be complex and chaotic, even when their basic physics includes only gravity and thermal pressure.  Gravity tends to create large density inhomogeneities, which in a disk can become sheared into spiral shapes or can develop into bound clumps, while the presence of coriolis forces and the possible role of tidal interactions provide additional ways for complex and unpredictable behavior to result.  Acoustic waves may also play a role in the dynamics of disks (Larson 1990; Bate \etal 2002a).  Clearly this extremely complex subject is still far from being fully understood, and more unexpected results may well emerge that make star and planet formation processes even more chaotic and unpredictable.  This might help to explain the great diversity of the planetary systems that have been discovered so far, but we may also have to accept that the formation of a planetary system closely similar to our own is a very improbable event. 

   A related problem that is still at an early stage of understanding is the origin of the bipolar jets that appear to be produced ubiquitously by newly formed stars.  The production of a jet clearly involves magnetic fields, and it is therefore a problem in magnetohydrodynamics.  Numerical work is beginning to provide useful insights by showing that when both rotation and a magnetic field are present, the result is quite generally the creation of a helically twisted magnetic field that tends to expel matter along the rotation axis (Banerjee and Pudritz 2006; Machida \etal 2006).  The associated magnetic torques may also remove significant angular momentum from protostellar disks and from their central stars, helping to drive accretion and to solve the angular momentum problem.  Thus it is possible that both gravitational and magnetic torques contribute to the solution of the angular momentum problem.  The detailed mechanism of jet production and the role of disks in this process are not yet well understood, but numerical simulations are increasingly playing a central role, and they will likely be indispensable for arriving at a better understanding of the origin of jets.

\section{Geometry, turbulence, and magnetic fields}

   In the formation of a system with more than one star, such as a binary or multiple system or a star cluster, the geometrical structure of a star-forming cloud plays a central role.  In general, complex non-spherical geometries can be expected.  Both prolate and oblate departures from spherical symmetry tend to grow with time in pressure-free collapse (Mestel 1965; Lin \etal 1965), and even with a finite pressure, prolate distortions can grow indefinitely to create very thin filaments (Larson 1972).  The similarity solutions discussed above are unstable to the growth of prolate distortions if $\gamma < 1.1$ (Hanawa and Matsumoto 2000), and simulations show that filamentary structures can result also from the collapse of a highly flattened cloud (Burkert and Hartmann 2004).  Rotation tends to produce oblate configurations which are often unstable to the development of bar-like deformations and to fragmentation into binary or multiple systems (Matsumoto and Hanawa 2003).  Spherical collapse is thus a very special case that is not likely ever to be closely realized; even if the initial configuration is roughly spherical, departures from spherical symmetry will tend to grow indefinitely as collapse proceeds.  A typical outcome of the collapse of a prestellar clump may be the formation of a binary or multiple system, as is indeed observed in all of the numerical simulations that have enough resolution to show binary formation (Larson 1978; Bate \etal 2002b.)

   Simulations of the collapse of large clouds show that they develop even more complex structures, including intricate networks of filaments (Klessen and Burkert 2001; Bonnell and Bate 2002; Bate \etal 2003; Jappsen \etal 2005; Klessen \etal 2007).  Networks of filaments interlaced with voids are found also in cosmological simulations (e.g., Springel \etal 2005a), and this suggests that filaments and filamentary networks may be a general feature of self-gravitating fluid systems.  Wispy and filamentary structures are indeed commonly observed in star-forming clouds (Schneider and Elmegreen 1979; Hartmann 2002).  As in cosmology, pressure is relatively unimportant on large scales in star-forming clouds, and the dominant effect generating structure in both cases may be the gravitational amplification of small density fluctuations in a medium where pressure is unimportant.  The strong tendency for filamentary structure to develop may result from the tendency noted above for prolate distortions to be amplified during collapse (Lin \etal 1965; Larson 1972).  While this effect could be expected to produce elongated structures, it was not anticipated before the simulations were done that the result would be such elaborate networks of filaments.  Thus numerical simulations have again revealed a new phenomenon, or a new level of complexity, that can play an important role in structuring star-forming clouds.  The development of filamentary networks is particularly well illustrated in cosmological simulations (e.g., Springel \etal 2005a), which presently have more detail than simulations of star formation because they have benefited from more computing power.  The fragmentation of filaments may play an important role in the origin of stellar masses, as will be discussed in section 6.

   The complex structures that appear in star-forming clouds typically contain many Jeans masses and therefore they form many stars, initially distributed along the filaments and clustered at the nodes of the networks (Jappsen \etal 2005; Klessen \etal 2007).  The stellar groupings that form tend to be hierarchical, consisting of smaller groups within larger ones (Bonnell \etal 2003), and they may even be fractal-like over some range of scales (Larson 1995; Elmegreen \etal 2000).  Hierarchical structure in star-forming clouds may result partly from a tendency for clouds to fragment into smaller units as they collapse, and partly from a tendency for smaller units to cluster gravitationally into larger ones.  Observations show, in any case, that newly formed stars are typically distributed in groups and clusters that are partly hierarchical in structure (Zinnecker \etal 1993; Larson 1995; Elmegreen \etal 2000; Testi \etal 2000).

   If stars form with a hierarchical spatial distribution, this can have important consequences for star formation processes because most stars then form close to other stars in dense groupings in which interactions can be important (Larson 2002; Bonnell \etal 2003).  Encounters with other forming stars in a dense environment may influence or even control the properties of binary systems and circumstellar disks by repeatedly disrupting them and forming new ones, as is seen to occur in some simulations (e.g., Bate \etal 2003).  Tidal interactions may also help to drive accretion in disks by creating strong disturbances that redistribute angular momentum and produce bursts of accretion onto the central star (Bonnell and Bastien 1992; Larson 2002).  In such environments, star formation must be a far more complex and chaotic process than that described by the idealized `standard model' of star formation (Shu 1977; Shu \etal 1987) which predicts a constant accretion rate.  Planet-forming disks may also experience frequent disturbances, and the planet formation process itself may be chaotic.  Evidence that our Solar system experienced an early disturbance is provided by the fact that its fundamental plane is tilted by 8 degrees from the solar rotation axis, plausibly because of an early encounter with another star in a dense environment (Heller 1993). 

   Fluid turbulence, or irregular fluid motions driven by hydrodynamic instabilities, may also have an important influence on the structure and dynamics of star-forming clouds (V\'azquez-Semadeni \etal 2000; Mac Low and Klessen 2004; Elmegreen and Scalo 2004; Ballesteros-Paredes \etal 2007).  Observations show that turbulence, or complex motions of some kind, nearly always supersonic, are ubiquitous in the interstellar medium and in interstellar clouds (Larson 1979, 1981).  Molecular clouds may then be understood as transient condensations in a turbulent medium, and the compression and shear processes associated with their formation may account for their often irregular and windblown appearances.  Simulations of supersonic turbulence show that wispy and filamentary structures are created quite generally, even when magnetic fields are present (Ostriker \etal 1999, 2001; Mac Low and Klessen 2004); turbulence may therefore create some of the filamentary structure seen in molecular clouds, and it may be responsible for some of their hierarchical character.  Another important result of simulations is that they show that turbulence is always highly dissipative, even when magnetic fields are important (Mac Low \etal 1998; Stone \etal 1998).  Contrary to what had previously been thought, the decay of turbulence is not retarded by a magnetic field, and this means that MHD turbulence cannot plausibly support star-forming clouds against collapse for more than a few crossing times.  This result removes some of the motivation for the magnetically supported quasi-static models of cloud evolution that had earlier been popular, and it is consistent with the view that star-forming clouds are transient and that star formation is a rapid process (Larson 1981, 1994; Elmegreen 2000; Hartmann 2003; Ballesteros-Paredes 2006).

   Magnetic fields thus do not appear to alter fundamentally the basic qualitative features and evolution of star-forming clouds (Ostriker \etal 1999, 2001; Mac Low and Klessen 2004), although they may tend to exert a damping effect on the dynamics and reduce the efficiency of star formation.  The basic qualitative aspects of star formation, such as the development of near-singularities in density, may still depend mainly on the interplay between gravity and thermal pressure, except for processes like jet production that clearly involve magnetic activity.  Magnetic fields may nevertheless influence some aspects of the geometry of star-forming clouds and star formation; for example, simulations of the collapse of magnetized clouds show that they tend to develop magnetized filaments and that gas flows tend to be channeled along flux tubes; protostellar collapse and accretion processes may then consist partly of flows along such flux tubes (Balsara \etal 2001; Banerjee \etal 2006).  The channeling of accretion flows along flux tubes might facilitate accretion and help to overcome stellar feedback effects.

\section{The origin of stellar masses}

   As we have seen, numerical simulations have done much to clarify the basic qualitative features of star formation, but an adequate understanding of the process must also account satisfactorily for the quantitative properties of the stars and stellar systems formed.  The most fundamental property of a star is its mass, and a central problem in star formation theory is understanding the origin of stellar masses, or the distribution of masses with which stars form.  The stars in nearby stellar systems seem to have formed with a nearly universal mass spectrum which peaks at a few tenths of a solar mass when given as the number of stars per unit log mass.  This typical stellar mass is of central importance in astronomy and astrophysics, and is the most important quantitative aspect of star formation needing to be explained.  The most familiar mass scale in star formation theory that might be relevant is the Jeans mass, which can be regarded as a dimensional mass obtained by balancing gravity and thermal pressure.  The relevance of the Jeans mass to star formation has been much debated, and other mass scales have also been proposed that depend on other physical effects such as turbulent or magnetic pressures (see the reviews by Larson 2003 and Bonnell \etal 2007).

   Detailed numerical simulations have allowed some of these competing ideas to be tested in a more realistic context than can be studied with analytic models.  Many simulations of cloud collapse and fragmentation have shown that, at least in the case of isothermal collapse without magnetic fields, the number of bound fragments formed is always of the same order as the number of Jeans masses initially present.  This result appears to be almost independent of the details of the initial conditions, and almost independent of the presence of turbulence (Larson 1978; Monaghan and Lattanzio 1991; Klessen and Burkert 2001; Klessen 2001; Bate \etal 2003; Bate and Bonnell 2005).  Thus, at least in the isothermal case, the Jeans mass clearly plays an important role in cloud fragmentation, even in situations that are much more complex and realistic than that originally considered by Jeans.  The details of the initial conditions appear to be largely forgotten as a cloud collapses, but the number of bound regions or clumps present nevertheless seems to be roughly conserved; some clumps may be dispersed and some may merge, but these attrition processes are roughly balanced by the continuing formation of new bound clumps.  The initial presence of turbulence in a cloud does not appear to make a large difference to the typical masses of the stars formed, although turbulence may influence various details such as the spatial distribution of these stars.

   The effect of magnetic fields on fragmentation is less well understood because simulations of fragmentation with magnetic fields are still at an early stage.  Existing results suggest that magnetic fields do not alter much the basic qualitative properties or evolution of turbulent star-forming clouds (Ostriker \etal 1999, 2001; Mac Low and Klessen 2004), and some results suggest that magnetic fields may not even change very much the typical masses of the bound clumps formed, which are still determined mainly by the requirement that gravity be able to overcome thermal pressure (P. S. Li \etal 2004; V\'azquez-Semadeni \etal 2005).  Detailed calculations of the collapse of magnetized cloud cores show that sufficient loss of magnetic flux occurs by ambipolar diffusion during the early stages of collapse that the later stages are largely governed by the same contest between gravity and thermal pressure that occurs in non-magnetic collapse (Basu and Mouschovias 1994; Ciolek and Basu 2000).  The Jeans mass may therefore still be relevant even in magnetized collapsing clouds.  Models of protostellar envelopes supported by `magnetic levitation' yield a mass scale in which magnetic pressure takes the place of gas pressure (Shu \etal 2004), but the role of this `magnetic Jeans mass' in cloud fragmentation remains to be clarified by simulations.  In any case, this magnetic Jeans mass is typically not very different from the thermal Jeans mass in dense cloud cores because the thermal and magnetic pressures are comparable there, as might perhaps be expected because gravitational instabilities can only grow if the gas pressure is at least equal to the magnetic pressure (Larson 1985).

   Simulations can now model the formation of clusters of a few hundred stars with at least moderate resolution, so they can study the resulting mass distribution and see how it depends on various physical effects and parameters.  Recent work has begun to address in more detail the thermal physics of collapsing clouds and the effect of departures from isothermality (Spaans and Silk 2000, 2005); the thermal physics might be expected to be important because the Jeans mass depends strongly on temperature, varying either as $T^{3/2}$ or as $T^2$ depending on whether the density or the pressure is specified.  An important result of recent collapse simulations, which again was not anticipated, is that if the usual isothermal equation of state is replaced with a polytropic one of the form $P \propto \rho^{\gamma}$, the amount of fragmentation that can occur is very sensitive to the exact value of $\gamma$ for values of $\gamma$ near unity.  For example, a simulation with $\gamma = 0.7$ produced about 380 bound fragments, while an otherwise identical one with $\gamma = 1.1$ produced only 18 fragments, a factor of 20 fewer (Y. Li \etal 2003).  Values of $\gamma$ between 0.7 and 1.1 are relevant in star-forming molecular clouds because at low densities the temperature is predicted to decrease with increasing density, following a trend that can be roughly approximated by $\gamma \simeq 0.73$, while at higher densities the temperature is predicted to rise slightly with increasing density, following approximately $\gamma \simeq 1.07$ (Larson 1985, 2005).

   If fragmentation occurs efficiently during the early stages of collapse when $\gamma \simeq 0.73$ but not during the later stages when $\gamma \simeq 1.07$, then the transition between these regimes where the temperature is a minimum might be expected to yield a preferred mass for fragmentation, since fragmentation to smaller masses should occur efficiently during the early low-density stages of collapse when $\gamma < 1$ but there should be relatively little fragmentation to yet smaller masses at the higher densities where $\gamma > 1$.  The Jeans mass at the transition point is predicted to be about 0.3 solar masses, which is in fact a typical stellar mass (Larson 1985, 2005).  Numerical experiments utilizing large simulations have shown that when a two-part polytropic equation of state is adopted that has $\gamma = 0.7$ at low densities and $\gamma = 1.1$ at high densities, the mass spectrum of the simulated stars peaks at a mass that is comparable to the Jeans mass at the critical density where the temperature is a minimum, and this peak mass scales in roughly the expected way when the critical density is varied (Jappsen \etal 2005; Bonnell \etal 2006).  These experiments confirm that the thermal behavior of collapsing clouds plays an important role in determining stellar masses, and they suggest that the Jeans mass at the density where the temperature is a minimum may be what primarily determines the typical stellar mass, although it cannot be excluded that other factors also play a role.

   The strong sensitivity of fragmentation to the value of $\gamma$ and to the transition from $\gamma < 1$ at low density to $\gamma > 1$ at high density can be understood if fragmentation occurs primarily in filaments, as is indeed the case in many of the simulations (Bonnell and Bate 2002; Jappsen \etal 2005; Klessen \etal 2007).  The isothermal case $\gamma = 1$ is a critical one for the collapse of a cylinder because if $\gamma < 1$, a cylinder can collapse indefinitely toward its axis and form an infinitely thin filament, whereas if $\gamma > 1$, the pressure in the cylinder rises faster than gravity and eventually halts collapse (Mestel 1965).  The intermediate case with $\gamma = 1$ is a marginally stable one.  These features are reflected in the properties of similarity solutions for collapsing cylinders that are analogous to the solutions described above for collapsing spheres; such solutions exist for cylinders if $\gamma < 1$ but not if $\gamma > 1$ (Kawachi and Hanawa 1998).  The collapse rate in the solutions with $\gamma < 1$ asymptotically approaches zero as $\gamma \rightarrow 1$, showing again that $\gamma = 1$ is a critical case.  Kawachi and Hanawa (1998) suggested that the deceleration of collapse that is expected to occur as $\gamma$ approaches unity will cause a collapsing filament to fragment into clumps at that point because the timescale for collapse toward the axis then becomes longer than the time required for the filament to fragment into clumps.  The geometry of star-forming clouds may therefore play an important role in the origin of stellar masses, along with the thermal physics, and the typical stellar mass of a few tenths of a solar mass may result from an interplay between the geometry and the thermal physics.

   In addition to the typical stellar mass, a second important feature of the stellar mass spectrum that needs to be explained is the declining power-law form at higher masses that was first found by Salpeter (1955).  The apparent near-universality of the Salpeter power law has inspired many theories to explain it, and there may be no single unique explanation (Bonnell \etal 2007).  It is worth noting, however, that the same simulations that reproduced the typical stellar mass also yielded without any extra physics an extension of the mass spectrum to high masses that resembles the Salpeter power law (Jappsen \etal 2005; Bonnell \etal 2006).  This suggests that the same basic elements of physics, i.e.\ gravity and thermal pressure, whose interplay yielded a characteristic mass may also result in a power-law extension of the mass spectrum to higher masses.  The detailed thermal physics does not seem to be so critical for the more massive stars, and similar results for their mass spectrum had been obtained earlier with an isothermal equation of state (Klessen 2001; Bonnell \etal 2001, 2003).  In all of these simulations, the most massive stars acquire most of their final mass by accretion from the surrounding cloud.  Accretion is more important for the more massive stars because the accretion rate increases strongly with mass, varying with the square of the mass for classical Bondi-Hoyle accretion (Bondi and Hoyle 1944; Bondi 1952).  The mass dependence of the accretion rate is further enhanced by the fact that the more massive stars tend to form in the dense central parts of forming clusters (see below).  For Bondi-Hoyle accretion, simple models predict an approach to a power-law mass spectrum that is somewhat shallower than the Salpeter function (Zinnecker 1982), but the steeper Salpeter slope can plausibly be recovered if the mass dependence of the accretion rate is increased by the tendency of the more massive stars to form in denser regions; however, there is no simple argument that yields exactly the Salpeter slope.  Nevertheless, the simulation results suggest that classical gravitational physics may be all that is needed to account for the power-law mass spectrum of the more massive stars.

   Numerical simulations have thus yielded two important insights concerning the origin of stellar masses: (1) the thermal physics of star-forming clouds can play an important role in determining typical stellar masses, and (2) continuing gas accretion in a cluster of forming stars can produce an extension of the mass spectrum to high masses that resembles the Salpeter power law.  These results were both suggested earlier on the basis of simple arguments and models, but the processes involved are sufficiently complex that detailed simulations are needed for quantitative predictions.  Much work remains to be done to explore the importance of these effects and others that may also be important, such as the chemical composition of star-forming clouds and the background radiation fields that are likely to be important in starburst regions and the early universe (Larson 2005).

\section{The formation of massive stars}

   An important but still poorly understood subject that has attracted much attention is the formation of the most massive stars (Garay and Lizano 1999; Stahler \etal 2000; Beuther \etal 2007).  Do these stars form by a scaled-up version of the processes that form low-mass stars, or in a qualitatively different way?  The formation of massive stars is almost certainly a more complicated process than the formation of low-mass stars because of the additional physical effects involved, including the effects of the strong radiation fields produced by these stars while they are still gaining mass.  For example, radiation pressure has been thought to pose an obstacle to continuing accretion and possibly to terminate accretion for masses greater than a few tens of solar masses (Larson and Starrfield 1971; Wolfire and Cassinelli 1987).  Partly for this reason, other mechanisms have been suggested for forming the most massive stars, including the possibility that they form not by gas accretion but by the merging of already formed less massive stars in a dense cluster core (Bonnell and Bate 2002; Bally and Zinnecker 2005).  Although a very high stellar density is required, mergers of young stars probably do occur at least occasionally because some stars are found in very close binary systems, and if nature can make stars in very close binary systems, it almost certainly makes at least a few that come close enough together to collide and merge.

   Detailed simulations of the effect of radiation pressure on the infalling envelopes of massive stars have shown, however, that radiation pressure does not stop accretion because radiation can readily escape in some directions while infall continues in others.  This can occur either via the formation of bipolar outflow cavities which allow radiation to escape along the cavity while gas continues to fall inward in other directions (Krumholz \etal 2005a), or via the formation of radiation-supported bubbles that expand outward while gas continues to flow in around them (Krumholz \etal 2005b).  While these results are perhaps not surprising in retrospect, they were not fully anticipated before the calculations were done, so this is another case where simulations have provided an important aid to imagination.  As a result of this work, it now appears that radiation pressure does not exclude any scenario involving the formation of massive stars by continuing accretion.  The importance of radiation pressure and other feedback effects may be further reduced if accretion flows are channeled along magnetized filaments (Balsara \etal 2001; Banerjee \etal 2006).

   Simulations of cluster formation show a tendency for the most massive stars to form near the center of a forming cluster where accretion and interaction rates are both greatly enhanced (Bonnell and Bate 2002; Bonnell \etal 2003).  This favors the building of massive stars by accretion in dense environments, but also allows the possibility that stellar collisions and mergers play a role.  Krumholz (2006) has noted that these simulations of cluster formation do not include the predicted radiative heating of the regions around forming massive stars, so that they overestimate the amount of fragmentation that can occur and hence the importance of interactions in these regions.  It has also been controversial whether `competitive accretion' is important in regions of massive star formation, i.e.\ whether different stars can accrete from a common gas reservoir (Krumholz \etal 2005c; Bonnell and Bate 2006).  Nevertheless, observations show that massive stars do often form in dense environments and in close proximity to other massive stars, and these observations, together with the fact that the simulations yield clusters resembling the observed ones with realistic stellar mass spectra, suggest that the simulations have at least some qualitative validity and that accretion and interactions play a role in the formation of massive stars.  Clearly much more work is needed on this very challenging problem, and detailed simulations including more physical effects will be needed to establish the importance of the various effects.

   As long as accretion plays any role in the formation of massive stars, the strong dependence of the accretion rate on mass makes this process a runaway one, approaching infinite mass within a finite time (Bonnell and Bate 2002).  The tendency of massive stars to form in the densest central part of a forming cluster further enhances the accretion rate and leads to the rapid building up of a high-mass tail on the mass spectrum (Bonnell \etal 2001, 2003).  The various dynamical processes occurring in the dense core of a forming cluster, including accretion, protostellar interactions, and the continuing condensation of matter into a smaller volume, all proceed faster as the density increases, making the cluster formation process itself a runaway one that simultaneously builds massive stars and a very compact and dense cluster core.  This process of coupled massive star and cluster formation (Bonnell \etal 2004) may be the real counterpart for massive stars of the development of a density singularity that occurs during the formation of low-mass stars.  The infalling envelope in the low-mass case is replaced in the high-mass case by a whole contracting cluster of forming stars whose dynamics is far more complex and chaotic.  If this picture is correct, it would represent a qualitative extension of our ideas about how stars form.  Gravity is still the driving force, and the process is still a highly inhomogeneous runaway one, but it is much more complex than before and produces an entire cluster of stars whose formation is coupled.  If no new mass scale is introduced, the result could be a scale-free mass spectrum similar to the Salpeter power law, as was indeed found in the simulations.

   Observations seem generally consistent with this picture.  One prediction that is confirmed by observations is that the mass of the most massive star in a young cluster is predicted to increase systematically with the mass of the cluster (Larson 2003; Weidner and Kroupa 2006).  According to the latter authors, the observed correlation is consistent with a picture in which stars of progressively larger mass form with a Salpeter-like mass distribution until star formation is terminated at a maximum mass when most of the gas has been consumed and the remaining gas is blown away by feedback effects.  This observed correlation places a strong constraint on models of massive star formation because it shows that massive stars know about the environment in which they form, and in particular that they know about the mass of the cluster in which they form.

\section{Gravitational processes on larger scales}

   Analogies almost certainly exist between the star formation processes discussed above and gravitational processes that occur on galactic and cosmological scales.  One such process is the formation of supermassive black holes at the centers of large galaxies; as is the case for the massive stars that form near the centers of clusters, the mass of the central black hole in a galaxy increases systematically with the mass of the galaxy, or at least with the mass of its bulge (Kormendy and Richstone 1995; Magorrian \etal 1998; Ferrarese and Merritt 2000; Gebhardt \etal 2000; Kormendy and Gebhardt 2001).  Because of this, the formation of central black holes in galaxies is believed to be closely coupled to the formation of the galaxies themselves, both being built up together by a series of mergers and accretion events.  Elaborate simulations have been made of galaxy mergers, some of which have included massive central black holes, and the results show that the black holes can experience strong gravitational drag effects and can fall rapidly toward the center of the merging system (Springel \etal 2005b; Di Matteo \etal 2005; Bekki \etal 2006).  In gas-rich mergers, the sinking of the black holes toward the center is accelerated by the gravitational drag produced by the gas, which itself falls rapidly toward the center.  When the black holes enter the nuclear region of the merged system, they can experience continuing gravitational drag by interaction with a nuclear gas disk, and this can cause them to merge in a time short compared with the time required for the galaxies to merge (Escala \etal 2005).  The processes involved in the formation and evolution of a close pair of orbiting black holes in a galactic nucleus may be similar to those involved in the formation of close binary systems of stars (Escala \etal 2003), as illustrated with lower resolution in the simulation by Bate \etal (2002b) of the formation of a small cluster of stars.

   Analogies very likely exist also between the hierarchical structure that develops in star-forming clouds and the hierarchical structure that appears on galactic and cosmological scales.  In both cases, simulations typically show the formation of networks of filaments, and matter tends to accumulate at the nodes of these networks to build systems of stars or galaxies.  The filamentary networks tend to be hierarchical, consisting of smaller structures within larger ones, and systems of larger and larger size are built up by the progressive merging of smaller units into larger ones.  The formation of aggregates of progressively larger size may be a general feature of the way gravity works in systems with structure on a range of scales, regardless of the origin of this structure, whether it be from primordial density fluctuations, gravitational instabilities, turbulence, or other phenomena.  Self-gravitating structures may generally be built or organized from the bottom up because gravitational processes operate faster in smaller and denser regions, so that matter is collected together first on small scales and then on progressively larger ones.  As smaller systems become assembled into progressively larger aggregates, the most massive individual objects in these systems, whether massive stars in clusters or supermassive black holes in galaxies, are built up at the same time by continuing gas accretion and possibly by occasional mergers with other similar objects.

   Simulations of galaxy formation show that the most dramatic events involved in this process are major mergers that rapidly bring a large amount of matter into a small volume; in particular, the gas in the merging systems quickly becomes highly concentrated into massive central clouds that produce intense starbursts and AGN activity (Springel \etal 2005b; Bekki \etal 2006).  Galaxies and their central black holes may both gain most of their mass during these events.  Analogous events may occur when star clusters are built up by the hierarchical merging of smaller groups of forming stars into larger ones, as occurs in several simulations of cluster formation (Bonnell and Bate 2002; Bonnell \etal 2003).  Mergers of groups of forming stars may be accompanied by bursts of accretion onto the forming stars, analogous to the bursts of black hole accretion that occur when galaxies merge; the merging of smaller groups into larger ones may then play an important role in the star formation process itself, especially the formation of the most massive stars.  The more detailed calculations that have been done for galaxy mergers that trigger bursts of black hole accretion (Springel \etal 2005b; Bekki \etal 2006) may therefore provide useful insights into the processes involved in the formation of massive stars.

   The creation of bound structures on a very wide range of scales, from individual objects to large aggregates, thus appears to be a general feature of gravitational condensation processes in both star formation and galaxy formation.  The processes of star formation and galaxy formation can therefore be modeled adequately only with simulations capable of incorporating a large range of scales and representing very complex dynamics.  Analytic models cannot adequately represent many of the essential features of star formation or galaxy formation because they cannot readily deal with the large range of scales involved or the complex and chaotic dynamics that are clearly intrinsic to these processes.

\section{Summary}

   Although the physics of star formation is classical and some of the basic concepts are old, numerical simulations have provided essential insights into how stars form in realistic situations that are not adequately described by analytical models.  They show that on the smallest scales, star formation begins with the runaway emergence of a nearly singular peak in density, surrounded by an infalling envelope that merges with the surrounding medium.  Forming stars are then predicted to gain most of their mass by accretion from their envelopes and from the surrounding medium.  Feedback effects may limit the efficiency of star formation by dispersing most of the gas in a star-forming cloud before it can condense into stars.  On the scale of clusters of stars, stars form in compact groups and multiple systems in which interactions can be important.  The gas being accreted by forming stars often forms disks around them, and gravitational instabilities in these disks or interactions with nearby stars may cause episodic bursts of accretion onto the central star.  Such instabilities or interactions may also perturb or disrupt forming planetary systems.  Magnetic fields may tend to channel gas flows along magnetized filaments, but may not otherwise change the basic qualitative features of star formation except for mediating the final stages of accretion and creating the bipolar jets that are frequently observed from newly formed stars.

   Simulations also show that star-forming clouds typically develop complex filamentary structures, which may be important for the origin of stellar masses because for filaments there is a preferred scale of fragmentation which is the Jeans mass at the density where the temperature is a minimum.  The formation of networks of filaments occurs in cosmological simulations as well, and appears to be a general feature of the evolution of self-gravitating fluid systems.  The appearance of hierarchical structure also appears to be a general feature, causing stars to form in groupings that are themselves initially hierarchical.  The formation of clusters of stars may then proceed like the formation of galaxies by the progressive merging of smaller groupings into larger ones.  In simulations of cluster formation, the most massive stars form near the centers of the forming clusters by continuing accretion and possibly also by mergers, processes possibly analogous to those involved in the formation of massive black holes at the centers of galaxies.  In both star formation and galaxy formation, structure is created over a very wide range of scales, and the dynamics is complex and chaotic.  Analytic models cannot adequately represent these processes, and detailed numerical simulations like those described in this article are therefore needed and can be expected to play an increasingly central role in advancing our understanding of such processes.

\References

\item[] Arce H G and Sargent A I 2006 {\it Astrophys.\ J.} {\bf 646} 1070--85

\item[] Balbus S A and Hawley J F 1998 {\it Rev.\ Mod.\ Phys.} {\bf 70}
  1--53

\item[] Ballesteros-Paredes J 2006 {\it Mon.\ Not.\ R. Astron.\ Soc.} {\bf 372}
  443--9

\item[] Ballesteros-Paredes J, Klessen R S, Mac Low M-M and V\'azquez-Semadeni
  E 2007 {\it Protostars and Planets V} ed B Reipurth \etal (Tucson:\ Univ.\
  of Arizona Press) pp~63--80

\item[] Bally J and Zinnecker H 2005 {\it Astron.\ J.} {\bf 129} 2281--93

\item[] Balsara D, Ward-Thompson D and Crutcher R M 2001 {\it Mon.\ Not.\ R.
  Astron.\ Soc.} {\bf 327} 715--20

\item[] Banerjee R and Pudritz R E 2006 {\it Astrophys.\ J.} {\bf 641} 949--60

\item[] Banerjee R, Pudritz R E and Anderson D W 2006 {\it Mon.\ Not.\ R.
  Astron.\ Soc.} {\bf 373} 1091--1106

\item[] Basu S 1997 {\it Astrophys.\ J.} {\bf 485} 240--53

\item[] Basu S and Mouschovias T Ch 1994 {\it Astrophys.\ J.} {\bf 432} 720--41

\item[] Bate M R, Bonnell I A 2005, {\it Mon.\ Not.\ R. Astron.\ Soc.}
  {\bf 356} 1201--21

\item[] Bate M R, Ogilvie G I, Lubow S H and Pringle J E 2002a {\it Mon.\ Not.\
  R. Astron.\ Soc.} {\bf 332} 575--600

\item[] Bate M R, Bonnell I A and Bromm V 2002b {\it Mon.\ Not.\ R. Astron.\
  Soc.} {\bf 336} 705--13

\item[] Bate M R, Bonnell I A and Bromm V 2003 {\it Mon.\ Not.\ R. Astron.\
  Soc.} {\bf 339} 577--99

\item[] Bekki K, Shioya Y and Whiting M 2006 {\it Mon.\ Not.\ R. Astron.\ Soc.}
  {\bf 371} 805--20

\item[] Beuther H, Churchwell E B, McKee C F and Tan J C 2007 {\it Protostars and
  Planets V} ed B Reipurth \etal (Tucson:\ Univ.\ of Arizona Press) pp~165--80

\item[] Black D C and Bodenheimer P 1976 {\it Astrophys.\ J.} {\bf 206} 138--49

\item[] Bodenheimer P 1995 {\it Ann.\ Rev.\ Astron.\ Astrophys.} {\bf 33}
  199--238

\item[] Bodenheimer P and Sweigart A 1968 {\it Astrophys.\ J.} {\bf 152}
  515--22

\item[] Bondi H 1952 {\it Mon.\ Not.\ R. Astron.\ Soc.} {\bf 112} 195--204

\item[] Bondi H and Hoyle F 1944 {\it Mon.\ Not.\ R. Astron.\ Soc.} {\bf 104}
  273--82

\item[] Bonnell I A and Bastien P 1992 {\it Astrophys.\ J.} {\bf 401} L31--4

\item[] Bonnell I A and Bate M R 2002 {\it Mon.\ Not.\ R. Astron.\ Soc.}
  {\bf 336} 659--69

\item[] Bonnell I A and Bate M R 2006 {\it Mon.\ Not.\ R. Astron.\ Soc.}
  {\bf 370} 488--94

\item[] Bonnell I A, Clarke C J, Bate M R and Pringle J E 2001 {\it Mon.\
  Not.\ R. Astron.\ Soc.} {\bf 324} 573--9

\item[] Bonnell I A, Bate M R and Vine S G 2003 {\it Mon.\ Not.\ R. Astron.\
  Soc.} {\bf 343} 413--8

\item[] Bonnell I A, Vine S G and Bate M R 2004 {\it Mon.\ Not.\ R. Astron.\
  Soc.} {\bf 349} 735--41

\item[] Bonnell I A, Clarke C J and Bate M R 2006 {\it Mon.\ Not.\ R.
  Astron.\ Soc.} {\bf 368} 1296--300

\item[] Bonnell I A, Larson R B and Zinnecker H 2007 {\it Protostars and
  Planets V} ed B Reipurth \etal (Tucson:\ Univ.\ of Arizona Press) pp~149--64

\item[] Burkert A and Hartmann L 2004 {\it Astrophys.\ J.} {\bf 616} 288--300

\item[] Ciolek G E and Basu S 2000 {\it Astrophys.\ J.} {\bf 529} 925--31

\item[] Di Matteo T, Springel V and Hernquist L 2005 {\it Nature} {\bf 433}
  604--7

\item[] Elmegreen B G 2000 {\it Astrophys.\ J.} {\bf 530} 277--81

\item[] Elmegreen B G and Scalo J 2004 {\it Ann.\ Rev.\ Astron.\ Astrophys.}
  {\bf 42} 211--73

\item[] Elmegreen B G, Efremov Y, Pudritz R E and Zinnecker H 2000
  {\it Protostars and Planets IV} ed V Mannings \etal (Tucson:\ Univ.\
  of Arizona Press) pp~179--215

\item[] Escala A, Larson R B, Coppi P S and Mardones D 2003 {\it Galactic Star
  Formation Across the Stellar Mass Spectrum (ASP Conf.\ Ser.\ Vol.\ 287)}
  ed J M De Buizer and N S van der Bliek (San Francisco: Astron.\ Soc.\ Pacific)
  pp~455--9

\item[] Escala A, Larson R B, Coppi P S and Mardones D 2005 {\it Astrophys.\
  J.} {\bf 630} 152--66

\item[] Ferrarese L and Merritt D 2000 {\it Astrophys.\ J.} {\bf 539} L9--12

\item[] Gammie C F 2001 {\it Astrophys.\ J.} {\bf 553} 174--83

\item[] Garay G and Lizano S 1999 {\it Publ.\ Astron.\ Soc.\ Pacific}
  {\bf 111} 1049--87

\item[] Gebhardt K, Bender R, Bower G, Dressler A, Faber S M, Filippenko A V,
  Green R, Grillmair C, Ho L C, Kormendy J, Lauer T R, Magorrian J, Pinkney J,
  Richstone D and Tremaine S 2000 {\it Astrophys.\ J.} {\bf 539} L13--16

\item[] Hanawa T and Matsumoto T 2000 {\it Publ.\ Astron.\ Soc.\ Japan}
  {\bf 52} 241--7

\item[] Hartmann L 1998 {\it Accretion Processes in Star Formation}
  (Cambridge: Cambridge Univ.\ Press)

\item[] Hartmann L 2002 {\it Astrophys.\ J.} {\bf 578} 914--24

\item[] Hartmann L 2003 {\it Astrophys.\ J.} {\bf 585} 398--405

\item[] Hartmann L and Kenyon S 1996 {\it Ann.\ Rev.\ Astron.\
  Astrophys.} {\bf 34} 207--40

\item[] Heller C H 1993 {\it Astrophys.\ J.} {\bf 408} 337--46

\item[] Hoyle F 1953 {\it Astrophys.\ J.} {\bf 118} 513--28

\item[] Hunter C 1977 {\it Astrophys.\ J.} {\bf 218} 834--45

\item[] Jappsen A-K, Klessen R S, Larson R B, Li Y and Mac Low M-M 2005
  {\it Astron.\ Astrophys.} {\bf 435} 611--23

\item[] Jeans J H 1902 {Phil.\ Trans.\ R. Soc.} {\bf 199} 49

\item[] Jeans J H 1929 {\it Astronomy and Cosmogony} (Cambridge:
  Cambridge Univ.\ Press; reprinted by Dover, New York, 1961)

\item[] Kawachi T and Hanawa T 1998 {\it Publ.\ Astron.\ Soc.\ Japan}
  {\bf 50} 577--86

\item[] Klessen R S 2001 {\it Astrophys.\ J.} {\bf 556} 837--46

\item[] Klessen R S and Burkert A 2001 {\it Astrophys.\ J.} {\bf 549}
  386--401

\item[] Klessen R S, Spaans M and Jappsen A-K 2007 {\it Mon.\ Not.\ R.
  Astron.\ Soc.} {\bf 374} L29--33

\item[] Kormendy J and Gebhardt K 2001 {\it 20th Texas Symposium on
  Relativistic Astrophysics} ed J C Wheeler and H Martel (AIP Conference
  Proceedings Vol.\ {\bf 586}, Springer, New York) pp~363--81

\item[] Kormendy J and Richstone D 1995 {\it Ann.\ Rev.\ Astron.\ Astrophys.}
  {\bf 33} 581--624

\item[] Krumholz M R 2006 {\it Astrophys.\ J.} {\bf 641} L45--8

\item[] Krumholz M R, McKee C F and Klein R I 2005a {\it Astrophys.\ J.}
  {\bf 618} L33--6

\item[] Krumholz M R, Klein R I and McKee C F 2005b {\it Massive Star Birth:
  A Crossroads of Astrophysics (IAU Symp.\ No.\ 227)} ed R Cesaroni \etal
  (Cambridge: Cambridge Univ.\ Press) pp~231--6

\item[] Krumholz M R, McKee C F and Klein R I 2005c {\it Nature} {\bf 438}
  332--4

\item[] Larson R B 1969 {\it Mon.\ Not.\ R. Astron.\ Soc.} {\bf 145} 271--95

\item[] Larson R B 1972 {\it Mon.\ Not.\ R. Astron.\ Soc.} {\bf 156} 437--58

\item[] Larson R B 1978 {\it Mon.\ Not.\ R. Astron.\ Soc.} {\bf 184} 69--85

\item[] Larson R B 1979 {\it Mon.\ Not.\ R. Astron.\ Soc.} {\bf 186} 479--90

\item[] Larson R B 1981 {\it Mon.\ Not.\ R. Astron.\ Soc.} {\bf 194} 809--26

\item[] Larson R B 1984 {\it Mon.\ Not.\ R. Astron.\ Soc.} {\bf 206} 197--207

\item[] Larson R B 1985 {\it Mon.\ Not.\ R. Astron.\ Soc.} {\bf 214} 379--98

\item[] Larson R B 1990 {\it Mon.\ Not.\ R. Astron.\ Soc.} {\bf 243} 588--92

\item[] Larson R B 1994 {\it The Structure and Content of Molecular Clouds
  (Lecture Notes in Physics No.\ 439)} ed T~L Wilson and K~J Johnston (Berlin:
  Springer) pp~13--28

\item[] Larson R B 1995 {\it Mon.\ Not.\ R. Astron.\ Soc.} {\bf 272} 213--20

\item[] Larson R B 2002 {\it Mon.\ Not.\ R. Astron.\ Soc.} {\bf 332} 155--64

\item[] Larson R B 2003 {\it Rep.\ Prog.\ Phys.} {\bf 66} 1651--1697

\item[] Larson R B 2005 {\it Mon.\ Not.\ R. Astron.\ Soc.} {\bf 359} 211--22

\item[] Larson R B and Starrfield S 1971 {\it Astron.\ Astrophys.} {\bf 13}
  190--97

\item[] Li P S, Norman M L, Mac Low M-M, Heitsch F 2004 {\it Astrophys.\ J.}
  {\bf 605} 800--18

\item[] Li Y, Klessen R S and Mac Low M-M 2003 {\it Astrophys.\ J.} {\bf 592}
  975--85

\item[] Lin C~C, Mestel L and Shu F H 1965 {\it Astrophys.\ J.} {\bf 142}
  1431--46

\item[] Lodato G and Rice W K M 2005 {\it Mon.\ Not.\ R. Astron.\ Soc.}
  {\bf 358} 1489--500

\item[] Low C and Lynden-Bell D 1976 {\it Mon.\ Not.\ R. Astron.\ Soc.}
  {\bf 176} 367--90

\item[] Mac Low M-M and Klessen R S 2004 {\it Rev.\ Mod.\ Phys.} {\bf 76}
  125--94

\item[] Mac Low M-M, Klessen R S, Burkert A and Smith M D 1998 {\it Phys.\
  Rev.\ Letters} {\bf 80} 2754--57

\item[] Machida M N, Inutsuka S-I and Matsumoto T 2006 {\it Astrophys.\ J.}
  {\bf 647} L151--4

\item[] Magorrian J, Tremaine S, Richstone D, Bender R, Bower G, Dressler A,
  Faber S M, Gebhardt K, Green R, Grillmair C, Kormendy J and Lauer T 1998
  {\it Astron.\ J.} {\bf 115} 2285--305

\item[] Matsumoto T and Hanawa T 2003 {\it Astrophys.\ J.} {\bf 595} 913--34

\item[] Matsumoto T, Hanawa T and Nakamura F 1997 {\it Astrophys.\ J.}
  {\bf 478} 569--84

\item[] Mestel L 1965 {\it Quart.\ J. R. Astron.\ Soc.} {\bf 6} 161--98

\item[] Monaghan J~J and Lattanzio J C 1991 {\it Astrophys.\ J.} {\bf 375}
  177--89

\item[] Nakamura F, Matsumoto T, Hanawa T and Tomisaka K 1999 {\it
  Astrophys.\ J.} {\bf 510} 274--90

\item[] Narita S, Hayashi C and Miyama S M 1984 {\it Prog.\ Theor.\ Phys.}
  {\bf 72} 1118--36

\item[] Norman M L, Wilson J R and Barton R T 1980 {\it Astrophys.\ J.}
  {\bf 239} 968--81

\item[] Ogino S, Tomisaka K and Nakamura F 1999 {\it Publ.\ Astron.\ Soc.\
  Japan} {\bf 51} 637--51

\item[] Ostriker E C, Gammie C F and Stone J M 1999 {\it Astrophys.\ J.}
  {\bf 513} 259--74

\item[] Ostriker E C, Stone J M and Gammie C F 2001 {\it Astrophys.\ J.}
  {\bf 546} 980--1005

\item[] Penston M V 1966 {\it Royal Observatory Bulletins, Series E, No.}
  {\bf 117} 299--312

\item[] Penston M V 1969a {\it Mon.\ Not.\ R. Astron.\ Soc.} {\bf 144} 425--48

\item[] Penston M V 1969b {\it Mon.\ Not.\ R. Astron.\ Soc.} {\bf 145} 457--85

\item[] Saigo K and Hanawa T 1998 {\it Astrophys.\ J.} {\bf 493} 342--50

\item[] Saigo K, Matsumoto T and Hanawa T 2000 {\it Astrophys.\ J.}
  {\bf 531} 971-87

\item[] Salpeter E~E 1955 {\it Astrophys.\ J.} {\bf 121} 161--7

\item[] Scalo J M 1987 {\it Interstellar Processes} ed D~J Hollenbach and
  H~A Thronson (Dordrecht:\ Reidel) pp~349--92

\item[] Schneider S and Elmegreen B G 1979 {\it Astrophys.\ J. Suppl.}
  {\bf 41} 87--95

\item[] Shu F H 1977 {\it Astrophys.\ J.} {\bf 214} 488--97

\item[] Shu F H, Adams F C and Lizano S 1987 {\it Ann.\ Rev.\ Astron.\
  Astrophys.} {\bf 25} 23--81

\item[] Shu F H, Li Z-Y and Allen A 2004 {\it Astrophys.\ J.} {\bf 601}
  930--51

\item[] Spaans M and Silk J 2000 {\it Astrophys.\ J.} {\bf 538} 115--20

\item[] Spaans M and Silk J 2005 {\it Astrophys.\ J.} {\bf 626} 644--8

\item[] Spitzer L 1978 {\it Physical Processes in the Interstellar
  Medium} (New York: Wiley-Interscience)

\item[] Springel V \etal 2005a {\it Nature} {\bf 435} 629--36

\item[] Springel V, Di Matteo T and Hernquist L 2005b {\it Mon.\ Not.\ R.
  Astron.\ Soc.} {\bf 361} 776--94

\item[] Stahler S W, Palla F and Ho P T P 2000 {\it Protostars and
  Planets IV} ed V Mannings \etal (Tucson:\ Univ.\ of Arizona Press)
  pp~327--51

\item[] Stone J M, Ostriker E C and Gammie C F 1998 {\it Astrophys.\ J.}
  {\bf 508} L99--102

\item[] Stone J M, Gammie C F, Balbus S A and Hawley J F 2000 {\it
  Protostars and Planets IV} ed V Mannings \etal (Tucson:\ Univ.\ of
  Arizona Press) pp~589--611

\item[] Testi L, Sargent A I, Olmi L and Onello J S 2000 {\it Astrophys.\ J.}
  {\bf 540} L53--6

\item[] Toomre A and Toomre J 1972 {\it Astrophys.\ J.} {\bf 178} 623--66

\item[] V\'azquez-Semadeni E, Ostriker E C, Passot T, Gammie C F and
  Stone J M 2000 {\it Protostars and Planets IV} ed V Mannings \etal
  (Tucson:\ Univ.\ of Arizona Press) pp~3--28

\item[] V\'azquez-Semadeni E, Kim J, Shadmehri M and Ballesteros-Paredes J
  2005 {\it Astrophys.\ J.} {\bf 618} 344--59

\item[] Vorobyov E I and Basu S 2006 {\it Astrophys.\ J.} {\bf 650} 956--69

\item[] Weidner C and Kroupa P 2006 {\it Mon.\ Not.\ R. Astron.\ Soc.}
  {\bf 365} 1333--47

\item[] Whitworth A and Summers D 1985 {\it Mon.\ Not.\ R. Astron.\ Soc.}
  {\bf 214} 1--25

\item[] Wolfire M G and Cassinelli J P 1987 {\it Astrophys.\ J.} {\bf 319}
  850--67

\item[] Zinnecker H 1982 {\it Symposium on the Orion Nebula to Honor Henry
  Draper} ed A~E Glassgold \etal (New York: Ann.\ New York Acad.\ Sci.\
  Vol.\ {\bf 395}) pp~226--35

\item[] Zinnecker H, McCaughrean M J and Wilking B A 1993 {\it
  Protostars and Planets III} ed E~H Levy and J~I Lunine (Tucson:
  Univ.\ of Arizona Press) pp~429--95

\endrefs

\end{document}